\documentclass[sigconf]{acmart}

\usepackage{multirow}
\usepackage{subfigure}
\usepackage{color}
\usepackage{colortbl}

\definecolor{c1}{RGB}{255,204,153}

\newcommand{\figref}[1]{Fig.~\ref{#1}}
\newcommand{\tabref}[1]{Table~\ref{#1}}
\newcommand{\equref}[1]{Eq.~(\ref{#1})}

\AtBeginDocument{%
  }

\copyrightyear{2024}
\acmYear{2024}
\setcopyright{acmlicensed}\acmConference[SIGGRAPH Conference Papers
'24]{Special Interest Group on Computer Graphics and Interactive Techniques
Conference Conference Papers '24}{July 27-August 1, 2024}{Denver, CO, USA}
\acmBooktitle{Special Interest Group on Computer Graphics and Interactive
Techniques Conference Conference Papers '24 (SIGGRAPH Conference Papers
'24), July 27-August 1, 2024, Denver, CO, USA}
\acmDOI{10.1145/3641519.3657462}
\acmISBN{979-8-4007-0525-0/24/07}

\begin{document}

\title{3D Gaussian Blendshapes for Head Avatar Animation}

\author{Shengjie Ma}
\orcid{0000-0001-9204-1624}
\affiliation{
    \institution{State Key Lab of CAD\&CG, Zhejiang University}
    \city{Hangzhou}
    \country{China}
}
\email{qtdysjj@gmail.com}

\author{Yanlin Weng}
\orcid{0000-0001-5223-4253}
\affiliation{
    \institution{State Key Lab of CAD\&CG, Zhejiang University}
    \city{Hangzhou}
    \country{China}
}
\email{weng@cad.zju.edu.cn}

\author{Tianjia Shao}
\orcid{0000-0001-5485-3752}
\affiliation{
    \institution{State Key Lab of CAD\&CG, Zhejiang University}
    \city{Hangzhou}
    \country{China}
}
\email{tjshao@zju.edu.cn}

\author{Kun Zhou}
\authornote{Corresponding author}
\orcid{0000-0003-4243-6112}
\affiliation{
    \institution{State Key Lab of CAD\&CG, Zhejiang University}
    \city{Hangzhou}
    \country{China}
}
\email{kunzhou@acm.org}

\renewcommand{\shortauthors}{Ma et al.}

\begin{abstract}
We introduce 3D Gaussian blendshapes for modeling photorealistic head avatars. Taking a monocular video as input, we learn a base head model of neutral expression, along with a group of expression blendshapes, each of which corresponds to a basis expression in classical parametric face models. Both the neutral model and expression blendshapes are represented as 3D Gaussians, which contain a few properties to depict the avatar appearance. The avatar model of an arbitrary expression can be effectively generated by combining the neutral model and expression blendshapes through linear blending of Gaussians with the expression coefficients. High-fidelity head avatar animations can be synthesized in real time using Gaussian splatting. Compared to state-of-the-art methods, our Gaussian blendshape representation better captures high-frequency details exhibited in input video, and achieves superior rendering performance.
\end{abstract}

\ccsdesc[500]{Computing methodologies~Reconstruction}
\ccsdesc[500]{Computing methodologies~Point-based models}

\keywords{Parametric face models, facial animation, facial tracking, facial reenactment}
\begin{teaserfigure} 
  \includegraphics[width=\textwidth]{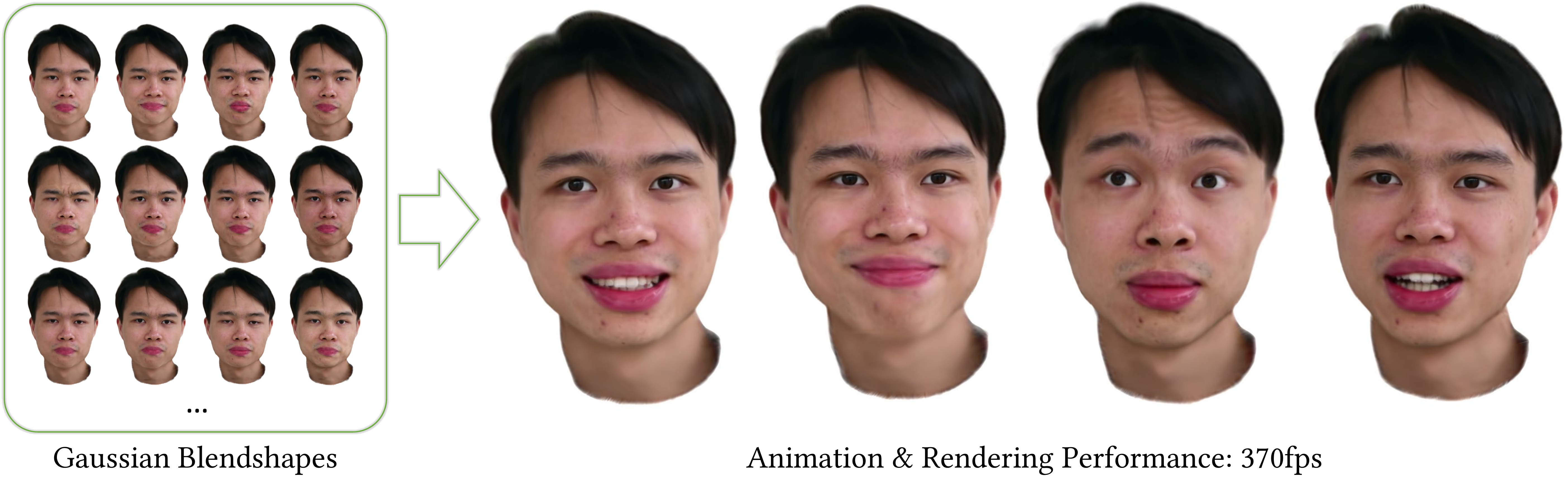}
  \caption{Our 3D Gaussian blendshapes are analogous to mesh blendshapes in classical parametric face models, which can be linearly blended with expressions coefficients to synthesize photo-realistic avatar animations in real time (370fps).}
  \Description{Four novel expressions on the right are generated by the Gaussian Blendshapes displayed in the left frame. The expressions include smiling, pursing lips, surprise, and talking.}
  \label{fig:teaser} 
\end{teaserfigure}

\maketitle

\section{Introduction}
Reconstructing and animating 3D human heads has been a long studied problem in computer graphics and computer vision, which is the key technology in a variety of applications such as telepresence, VR/AR and movies.
Most recently, head avatars based on neural radiance fields (NeRF)~\cite{NeRF20} demonstrate great potential in synthesizing photorealistic images. These techniques achieve dynamic avatar control typically by conditioning NeRFs on a parametric head model~\cite{zielonka2023instant} or expression codes~\cite{gafni2021dynamic}. Gao et al.~\shortcite{gao2022reconstructing} and Zheng et al.~\shortcite{zheng2022avatar} instead propose to construct a set of NeRF blendshapes and linearly blend them to animate the avatar.

The blendshape model is a classic representation for avatar animation. It consists of a group of 3D meshes, each of which corresponds to a basis expression. The face shape of an arbitrary expression can be efficiently computed by linearly blending the basis meshes with corresponding expression coefficients. The advantages of easy-to-control and high efficiency make blendshape models the most popular representation in professional animation production~\cite{LewisEG2014} as well as consumer avatar applications (e.g., iPhone Memoji)~\cite{WENG2014172}.

In this paper, we introduce a 3D Gaussian blendshape representation for constructing and animating head avatars. We build the representation upon 3D Gaussian splatting (3DGS)~\cite{3DGS}, which represents the radiance field of a static scene as 3D Gaussians and provides compelling quality and speed in novel view synthesis. Our representation consists of a base model of neutral expression and a group of expression blendshapes, all represented as 3D Gaussians. Each Gaussian contains a few properties (e.g., position, rotation and colors) as in 3DGS and depicts the appearance of the head avatar. Each Gaussian blendshape corresponds to a mesh blendshape of traditional parametric face models~\cite{Facewarehouse14,FLAME17} and has the same semantic meaning. A Gaussian head model of an arbitrary expression can be generated by blending the Gaussian blendshapes with the expression coefficients, which is rendered to high-fidelity images in real time using Gaussian splatting. The motion parameters tracked by previous face tracking algorithms (e.g., \cite{CaoDDE2014,Zielonka2022TowardsMR}) can be used to drive the Gaussian blendshapes to produce head avatar animations.

We propose to learn the Gaussian blendshape representation from a monocular video. We use previous methods to construct the mesh blendshapes from the input video, and distribute a number of Gaussians on the mesh surfaces as an initialization. We then jointly optimize all Gaussian properties. As Gaussian blendshapes are driven by the same expression coefficients for mesh blendshapes, each Gaussian blendshape must be semantically consistent with its corresponding mesh blendshape, i.e., the differences between the Gaussian blendshape and neutral model should be consistent with the differences between the corresponding mesh blendshape and neutral mesh.
Directly optimizing Gaussian properties without considering blendshape consistency causes overfitting and artifacts for novel expressions unseen in training.  
To this end, we present an effective strategy to guide the Gaussian optimization to follow the consistency requirement. Specifically, we introduce an intermediate variable to formulate the Gaussian difference as terms proportional to the mesh difference. By optimizing this intermediate variable directly during training, we produce Gaussian blendshapes differing from the neutral model in a consistent way that mesh blendshapes differ from the neutral mesh.

Extensive experiments demonstrate that our Gaussian blendshape method outperforms state-of-the-art methods~\cite{zheng2023pointavatar,zielonka2023instant,gao2022reconstructing} in synthesizing high-fidelity head avatar animations that best capture high-frequency details observed in input video, and achieving significantly faster speeds in avatar animation and rendering (see \figref{fig:teaser}).

\section{Related work} 

Researchers have proposed various representations for head avatars. Early works employ explicit 3D mesh representation to reconstruct the 3D shape and appearance from images. The seminal work~\cite{BlanzV99} proposes the 3D Morphable Model (3DMM) to model the face shape and texture on a low-dimensional linear subspace. There are many follow-up works along this direction such as full-head models~\cite{PloumpisVSMWPSG21}, and deep non-linear models~\cite{Tran018}. The 3D mesh representation is also used to build riggable heads for head animation~\cite{HuSWNSFSSCL17,BaiC0T21,ChaudhuriVSW20}. To generate detailed animations, researchers further propose image-based dynamic avatars controlling the full head with hair and headwear~\cite{CaoWWSZ16}, or additionally reconstruct fine-level correctives~\cite{CVVPT16, FengFBB21,IchimBP15,Yang0WHSYC20}.

In order to achieve high realism rendering, recent approaches utilize neural radiance fields (NeRF)~\cite{NeRF20} to implicitly represent head avatars and have achieved impressive results~\cite{gafni2021dynamic,xu2023latentavatar,grassal2022neural,lombardi2021mixture, zheng2022avatar, xu2022surface,xu2023avatarmav,Jiang0BZ22}. For instance, i3DMM~\cite{YenamandraTBSEC21} presents the first neural implicit function based on the 3D morphable model of full heads.
HeadNerf~\cite{hong2022headnerf} introduces a NeRF-based parametric head model that integrates the neural radiance field to the parametric representation of the head.
The state-of-the-art work INSTA~\cite{zielonka2023instant} models a dynamic neural radiance field based on InstantNGP~\cite{muller2022instant} embedded around a parametric face model. It is able to reconstruct a head avatar in less than 10 minutes. PointAvatar~\cite{zheng2023pointavatar} presents a point-based representation and learns a deformation field based on FLAME's expression vectors to drive the points. 
NeRFBlendshape~\cite{gao2022reconstructing} constructs NeRF-based blendshape models for semantic animation control and photorealistic rendering by combining multi-level voxel fields with expression coefficients.

\begin{figure*}[t]
\centering
\includegraphics[width=\linewidth]{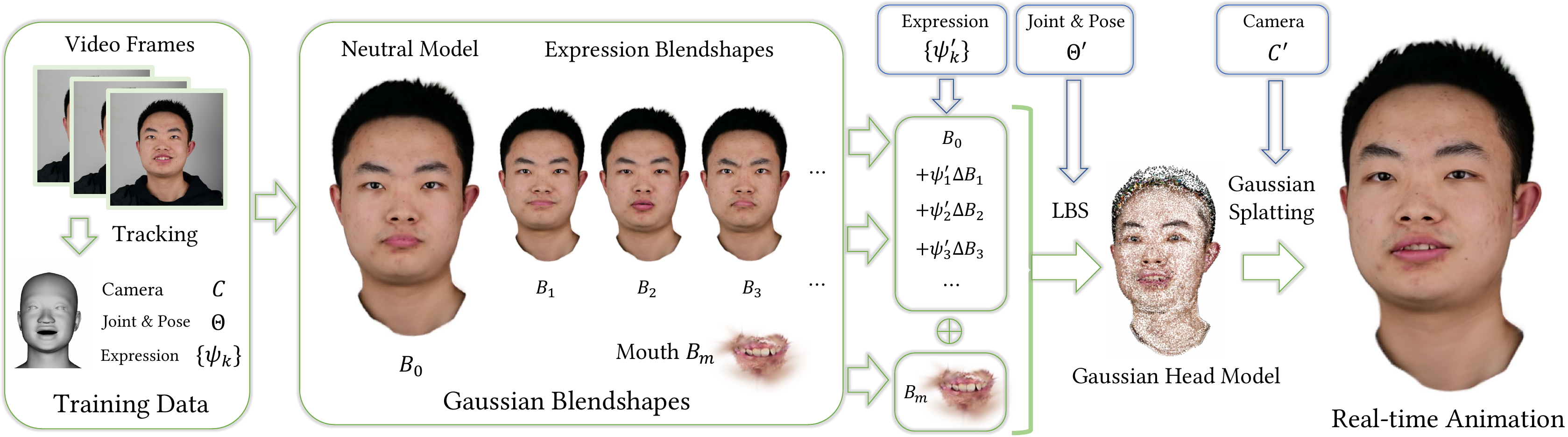}
\caption{Overview of our method. Taking a monocular video as input, our method learns a Gaussian blendshape representation of a head avatar, which consists of a neutral model $B_0$, a group of expression blendshapes $\{B_1,B_2,...,B_K\}$, and the mouth interior model $B_m$, all represented as 3D Gaussians. Avatar models of arbitrary expressions and poses can be generated by linear blending with expression coefficients $\{\psi'_k\}$ and linear blend skinning with joint and pose parameters $\Theta'$, from which we render high-fidelity images in real time using Gaussian splatting.}
\Description{A series of block diagrams, arranged from left to right, illustrate the pipeline. The first block contains the input video and the tracking parameters. The second block shows the optimized Gaussian blendshapes. The final block depicts the compositing and splatting processes, which utilize Gaussian blendshapes to render novel images.}
\label{fig:pipeline}
\end{figure*}

Many concurrent works have been proposed to apply the 3D Gaussian representation introduced by \cite{3DGS} to construct head avatars~(e.g., \cite{qian2023gaussianavatars,chen2023monogaussianavatar,xu2023gaussian,dhamo2023headgas,wang2023gaussianhead,saito2023relightable,xiang2023flashavatar}). Most of them use the 3D Gaussian representation together with neural networks. For example, GaussianHead~\cite{wang2023gaussianhead} uses Multi-layer Perceptrons (MLPs) to decode the dynamic geometry and radiance parameters of Gaussians. FlashAvatar~\cite{xiang2023flashavatar} attaches Gaussians on a mesh with learnable offsets, which are represented as MLPs. Saito et al.~\shortcite{saito2023relightable} construct relightable head avatars by using networks to decode the parameters of 3D Gaussians and learnable radiance transfer functions. To our knowledge, none of concurrent works introduce the idea of Gaussian blendshapes as in our paper. A unique advantage of our method is that it  only requires linear blending of Gaussian blendshapes to construct a head avatar of arbitrary expressions, which brings significant benefits in both training and runtime performance. The method closest to our work in terms of performance is FlashAvatar~\cite{xiang2023flashavatar}, which achieves 300fps for 10k Gaussians and degrades to $\sim$100fps for 50k Gaussians, while we achieve 370fps for 70k Gaussians.

\section{Method}

\subsection{3D Gaussian BlendShapes}
\label{sec:3dgb}
Our Gaussian blendshape representation consists of a neutral base model $B_0$ and a group of expression blendshapes $\{B_1,B_2,...,B_K\}$, all represented as a set of 3D Gaussians, each of which has a few basic properties (i.e., position $\mathbf{x}$, opacity $\alpha$, rotation $\mathbf{q}$, scale $\mathbf{s}$ and spherical harmonics coefficients $SH$) as in 3DGS~\cite{3DGS}. Each Gaussian of $B_0$ also has a set of blend weights  $\mathbf{w}$ for joint and pose control. There is a one-to-one correspondence between the Gaussians of $B_0$ and each blendshape $B_k$. The deviation of $B_k$ from $B_0$ can be defined as the difference between their Gaussian properties, $\Delta B_k = B_k - B_0$. The head avatar model of an arbitrary expression is computed as:
\begin{equation}\label{eq:avatar}
    B^\psi = B_0 + \sum_{k=1}^{K}{\psi}_k\Delta B_k,
\end{equation}
where $\{{\psi}_k\}$ are the expression coefficients.

Currently we use the Principal Component Analysis (PCA) based blendshape model FLAME~\cite{FLAME17}, although other muscle-inspired blendshapes such as the Facial Action Coding System (FACS) based model FaceWarehouse~\cite{Facewarehouse14} can be also employed.
Besides facial expression control, FLAME also provides joint and pose parameters, $\Theta$, for controlling the motions of head, jaw, eyeballs and eyelids, which are used with linear blend skinning (LBS) to transform the head avatar model (i.e., its Gaussians): $B^{\psi*} = LBS(B^\psi,\Theta)$, where the blend weights associated with Gaussians of $B_0$ are used.

\paragraph{Mouth Interior Gaussians.} The motions of mouth interior and hair are usually not affected by facial expressions, and thus not covered in the FLAME mesh, neither the blendshape model described above. Hair can move with the head rigidly, while the motion of teeth is controlled by the jaw joint in FLAME. We find that in practice the blendshape Gaussians generated in our training are able to model hair well, but the mouth interior results are not good enough. We thus define a separate set of Gaussians for mouth interior, $B_m$, which move with the jaw joint in FLAME. The properties of these mouth Gaussians do not change with expressions, but are only transformed with the jaw joint, i.e., $B_m^* = LBS(B_m,\Theta)$.

Finally, the transformed Gaussian model ($B^{\psi*}$, $B_m^*$) can be rendered to high-fidelity images $I_{r}$ in real time using Gaussian splatting. \figref{fig:pipeline} shows the overview of our method.

\subsection{Training}
\paragraph{Data Preparation.} 
Following \cite{zielonka2023instant}, we use the face tracker of \cite{Zielonka2022TowardsMR} to compute the FLAME meshes of neutral expression and $K=50$ basis expressions, as well as the camera parameters, joint and pose parameters, and expression coefficients, for each video frame. We also extract the foreground head mask for each input frame. 

\paragraph{Initialization.} 
We first initialize the neutral model $B_0$, expression blendshapes $\{B_k\}$, as well as the mouth interior Gaussians $B_m$.
For $B_0$, we distribute a number of points on the neutral FLAME mesh $M_0$ using Poisson disk sampling~\cite{BowersWWM10}, and use them as the initialization of Gaussian positions. Other Gaussian properties are initialized as in 3DGS~\cite{3DGS}. For each Gaussian, we also find its closest triangle on $M_0$, and compute its LBS blend weights as the linear interpolation of blend weights of the triangle vertices. To initialize the mouth interior Gaussians $B_m$, we use two pre-defined billboards to represent the upper and lower teeth, which are sampled to Gaussians using Poisson disk sampling. The upper teeth Gaussians are rigidly bound to the back of the head, while lower teeth Gaussians are bound to the vertex having the largest skinning weight for the jaw joint.

To initialize the expression blendshape $B_k$, we transform each Gaussian of $B_0$ using the deformation gradients~\cite{sumner2004deformation} from $M_0$ to the expression FLAME mesh $M_k$. Specifically, for each neutral Gaussian $G^i_0$, we compute the affine transformation from its closest triangle on $M_0$ to the corresponding triangle on $M_k$, and extract the rotation component~\cite{shoemake1992matrix}, which is applied to the position, rotation and spherical harmonics (SH) coefficients of $G^i_0$ to yield the corresponding Gaussian $G^i_k$ of expression blendshape $B_k$. Note that we omit the scale component as we find the transformation is very close to rigid. The scale and opacity properties of $G^i_k$ are kept the same as those of $G^i_0$. In this way, we can construct each expression blendshape $B_k$ from $B_0$, as well as their difference $\Delta B_k = B_k - B_0$. 

\paragraph{Optimization.} 
After initialization, we jointly optimize $B_0$, $\{\Delta B_k\}$, and $B_m$. For each video frame, we reconstruct the Gaussian head model $B_{\psi}$ by linearly blending $B_0$ and $\{\Delta B_k\}$ with the tracked expression coefficients according to \equref{eq:avatar}, and then transform $B_{\psi}$ and $B_m$ using LBS with the tracked joint and pose parameters: $B^{\psi*} = LBS(B^\psi,\Theta)$, $B_m^* = LBS(B_m,\Theta)$. Finally, we get the rendered image from $B^{\psi*}$ and $B_m^*$ using Gaussian splatting. The optimization process is similar to 3DGS~\cite{3DGS}, which also involves adaptive density control steps of adding and removing Gaussians. 

\begin{figure}[h]
\centering
\includegraphics[width=\linewidth]{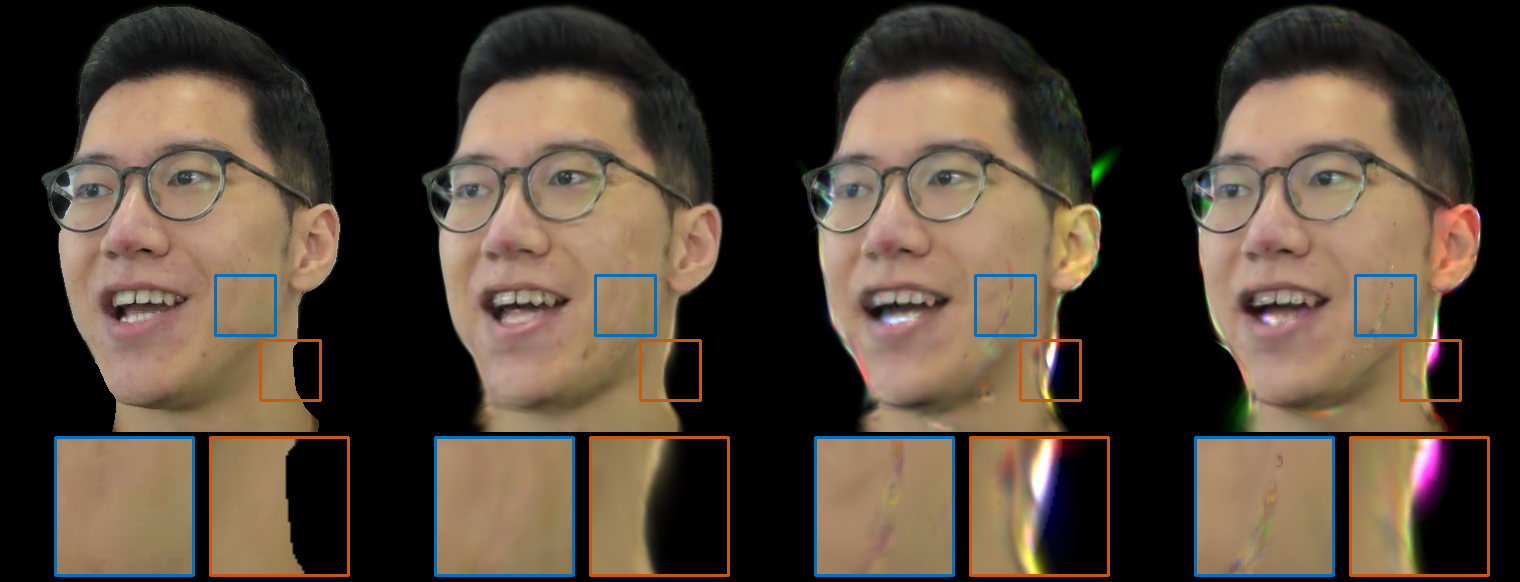}
\makebox[0.24\linewidth]{Ground truth}
\makebox[0.24\linewidth]{w/ consistency}
\makebox[0.24\linewidth]{w/ position}
\makebox[0.24\linewidth]{w/o consistency}
\makebox[0.24\linewidth]{}
\makebox[0.24\linewidth]{(ours)}
\makebox[0.24\linewidth]{consistency only}
\makebox[0.24\linewidth]{}
\caption{Ablation study on blendshape consistency. The optimization without blendshape consistency leads to apparent artifacts like dirty color and glitch in both interior and boundary areas. Enforcing blendshape consistency only on Gaussian positions also leads to poor results.} 
\Description{Figure 4. Fully described in the text.}
\label{fig:ablation_artifact}
\end{figure}

During optimization, a crucial thing to avoid overfitting is to preserve the semantic consistency between each Gaussian blendshape $B_k$ and its corresponding mesh blendshape $M_k$. As aforementioned, the Gaussian blendshapes are blended using the same tracked expression coefficients based on the parametric mesh model of FLAME, in both training and runtime computations. To ensure the semantic validity of such blending calculation, the difference between $B_k$ and $B_0$ (i.e., $\Delta B_k$) must be consistent with the difference between $M_k$ and $M_0$ (i.e., $\Delta M_k$), which means in head regions having large vertex position differences between $M_k$ and $M_0$, the Gaussian differences between $B_k$ and $B_0$ should also be large, and small otherwise. Directly optimizing $\{\Delta B_k\}$ without such consistency consideration will lead to overfitting, where apparent artifacts easily occur on novel expression coefficients unseen in the training images (see \figref{fig:ablation_artifact} for examples). 

However, unlike $\Delta M_k$ only containing vertex position displacements, $\Delta B_k$ contains different kinds of properties, such as position, rotation, and color. It is thus difficult to design a loss function term to explicitly enforce consistency between $\Delta B_k$ and $\Delta M_k$, while not sacrificing the image loss. Instead, we propose a simple yet effective strategy to guide the Gaussian optimization to implicitly follow the consistency requirement. Specifically, for each Gaussian $G_i$, let $\Delta G_{i,k}$ be the difference between its properties in $B_k$ and $B_0$.  
We introduce an intermediate variable, $\Delta \widehat{G}_{i,k}$, to formulate $\Delta G_{i,k}$ as:
\begin{equation}
\label{eq:dif_scale}
    \Delta G_{i,k} = \Delta G^{init}_{i,k} + max(f(d_{i,k}),0) \Delta \widehat{G}_{i,k},
\end{equation}
where $\Delta G^{init}_{i,k}$ is the initial value of $\Delta G_{i,k}$ calculated in the aforementioned initialization stage and regarded as a constant during optimization, and $d_{i,k}$ is the magnitude of position displacement of the surface point closest to $G_i$, from $M_0$ to $M_k$. The linear function $f(x)=(x-\epsilon)/(\tilde{d}-\epsilon)$ scales the maximum magnitude of position difference $\tilde{d}$ between $M_k$ and $M_0$ to 1, and a threshold magnitude $\epsilon=0.00001$ to 0. The $max$ function is necessary to avoid negative scaling values for positional displacement magnitudes below $\epsilon$.

\begin{table*}[t]
\caption{Quantitative comparisons between INSTA~\cite{zielonka2023instant}, PointAvatar~\cite{zheng2023pointavatar}, and our method.}
\label{tab:evaluation_INSTA}
\centering
\tabcolsep=0.15cm
\resizebox{\linewidth}{!}{
\begin{tabular}{|cc|cccccccccccc}
\hline
\multicolumn{2}{c|}{\multirow{2}{*}{Datasets}} & \multicolumn{8}{c|}{INSTA dataset} & \multicolumn{4}{c}{Our dataset} \\  \cline{3-14} \multicolumn{2}{c|}{} & bala & biden & justin & malte\_1 & marcel & nf\_01 & nf\_03 &\multicolumn{1}{c|} {wojtek\_1} & subject1 & subject2 & subject3 & subject4  \\ \hline
\multicolumn{1}{c|}{\multirow{3}{*}{PSNR $\uparrow$}}  & INSTA   & 28.66  & 28.38  & 29.74 & 26.27 & 23.75  & 25.89 & 26.10  & \multicolumn{1}{c|}{29.84}    & 28.88 & 28.16 & 28.60  & 30.83 \\ 
\multicolumn{1}{c|}{}                       & PointAvatar   & 29.60 & 31.72 & 32.31 & 27.46 & 24.60 & \cellcolor{c1}28.34 & \cellcolor{c1}29.82 & \multicolumn{1}{c|} {31.94} & 31.43 & \cellcolor{c1}32.57 & 30.95 & 32.57 \\ 
\multicolumn{1}{c|}{}                       & Ours & \cellcolor{c1}33.34 & \cellcolor{c1}32.48 & \cellcolor{c1}32.49 & \cellcolor{c1}28.56 & \cellcolor{c1}26.61 & 27.92 & 28.62 & \multicolumn{1}{c|}{\cellcolor{c1}32.39} & \cellcolor{c1}32.87 & 32.55 & \cellcolor{c1}31.54 & \cellcolor{c1}34.03 \\ \cline{1-14}
\multicolumn{1}{c|}{\multirow{3}{*}{SSIM $\uparrow$}}  & INSTA  & 0.9130 & 0.9484 & 0.9530 & 0.9262 & 0.9133 & 0.9246 & 0.9129 & \multicolumn{1}{c|}{0.9457} & 0.9195 & 0.9443 & 0.9078 & 0.9445 \\ 
\multicolumn{1}{c|}{}                       & PointAvatar   & 0.9099 & 0.9565 & 0.9595 & 0.9225 & 0.9121 & 0.9278 & 0.9208 & \multicolumn{1}{c|}{0.9502} & 0.9219 & 0.9463 & 0.9062 & 0.9433 \\ 
\multicolumn{1}{c|}{}                       & Ours & \cellcolor{c1}0.9490 & \cellcolor{c1}0.9672 & \cellcolor{c1}0.9696 & \cellcolor{c1}0.9461 & \cellcolor{c1}0.9348 & \cellcolor{c1}0.9448 & \cellcolor{c1}0.9381 & \multicolumn{1}{c|}{\cellcolor{c1}0.9645} & \cellcolor{c1}0.9384 & \cellcolor{c1}0.9577 & \cellcolor{c1}0.9268 & \cellcolor{c1}0.9646 \\ \cline{1-14}
\multicolumn{1}{c|}{\multirow{3}{*}{LPIPS $\downarrow$}} & INSTA & 0.0817 & 0.0545 & \cellcolor{c1}0.0614 & 0.0751 & 0.1540 & 0.1285 & 0.1137 & \multicolumn{1}{c|}{\cellcolor{c1}0.0588} & 0.1536 & 0.1208 & 0.1733 & 0.1144 \\ 
\multicolumn{1}{c|}{}                       & PointAvatar & 0.0821 & 0.0535 & 0.0649 & 0.0718 & 0.1574 & 0.1350 & 0.1221 & \multicolumn{1}{c|}{0.0661} & 0.1568 & \cellcolor{c1}0.1190 & 0.1715 & 0.1285 \\ 
\multicolumn{1}{c|}{}                       & Ours & \cellcolor{c1}0.0772 & \cellcolor{c1}0.0522 & 0.0631 & \cellcolor{c1}0.0703 & \cellcolor{c1}0.1391 & \cellcolor{c1}0.1174 & \cellcolor{c1}0.0965 & \multicolumn{1}{c|} {0.0595} & \cellcolor{c1}0.1520 & 0.1197 & \cellcolor{c1}0.1689 & \cellcolor{c1}0.1075 \\ \hline
\end{tabular}
}
\end{table*}

\begin{figure}[t]
\setlength{\columnsep}{0pt}
  \begin{minipage}[c]{\dimexpr0.03\linewidth}
    \rotatebox{90}{
    \begin{tabular}{p{2.4cm}p{2.6cm}p{1.8cm}}
        w/o consistency & w/ consistency & Mesh disp. 
    \end{tabular}
    }
  \end{minipage}
  \begin{minipage}[c]{\dimexpr0.95\linewidth-\columnsep}
    \includegraphics[width=\linewidth]{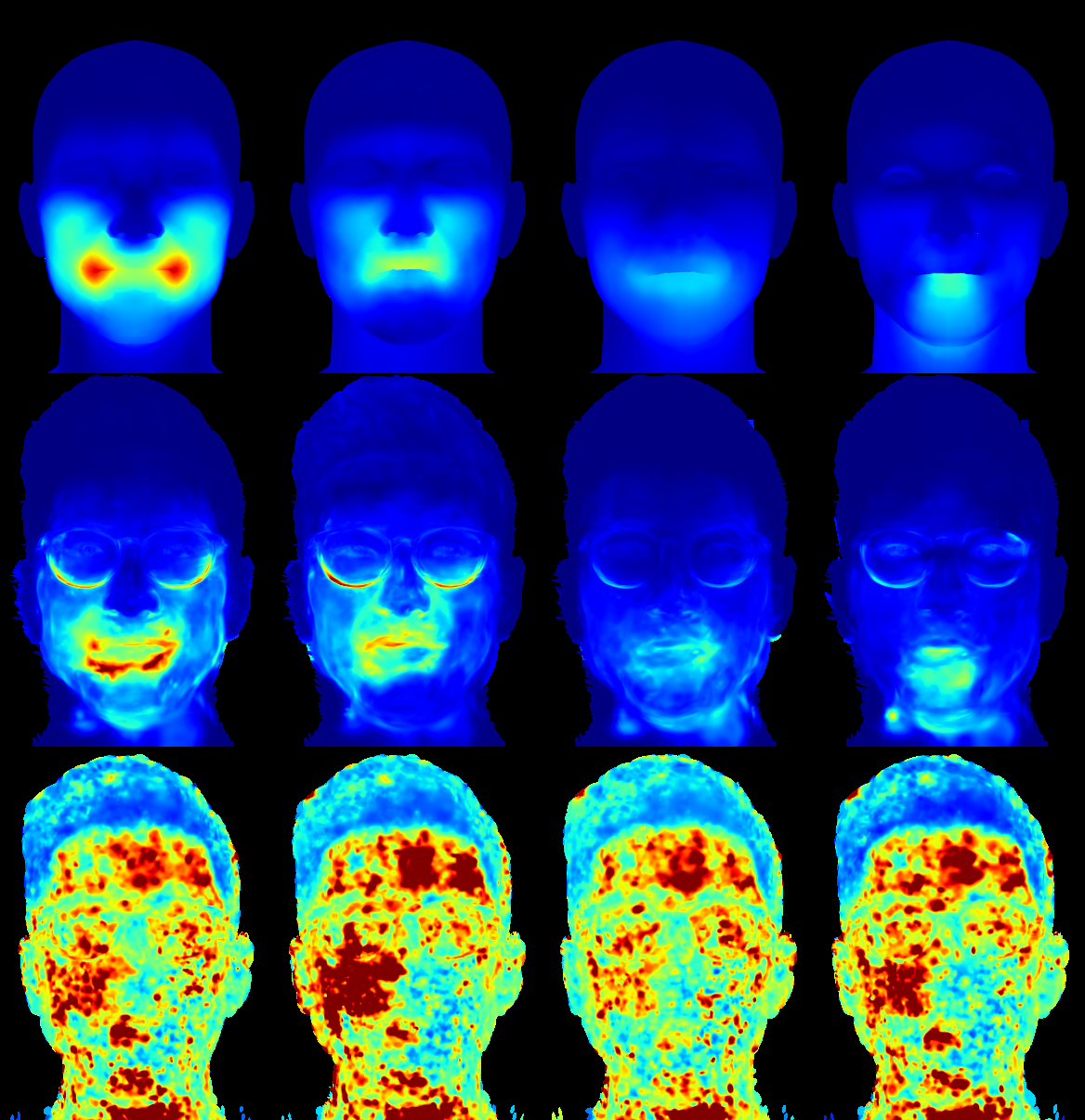}
  \end{minipage}
\makebox[0.03\linewidth]{}
\makebox[0.23\linewidth]{Blendshape1}
\makebox[0.23\linewidth]{Blendshape2}
\makebox[0.23\linewidth]{Blendshape3}
\makebox[0.23\linewidth]{Blendshape4}
\caption{The impact of the blendshape consistency on the optimization of expression blendshapes. The first row shows the displacement magnitude between $M_k$ and $M_0$. The second and the third rows show the magnitude of optimized $\Delta B_k$ with or without blendshape consistency. }
\Description{The differences between each individual blendshape and the base model are visualized. Gaussian blendshapes considering blendshape consistency in optimization exhibit a pattern similar to mesh blendshapes, while those without blendshape consistency are quite different.}
\label{fig:prove_consistency}
\end{figure}

\equref{eq:dif_scale} essentially represents the actual Gaussian difference $\Delta G_{i,k}$ as the sum of its initial value and the scaled value of $\Delta \widehat{G}_{i,k}$ according to its corresponding positional displacement in mesh blendshapes, which effectively correlates Gaussian differences with position displacements. The initial Gaussian difference $\Delta G^{init}_{i,k}$ is proportional to the position displacement of mesh blendshapes, as it is computed using the deformation gradients from $M_0$ to $M_k$. Scaling $\Delta \widehat{G}_{i,k}$ according to positional displacement ensures that the Gaussian difference $\Delta G_{i,k}$ is updated at a rate proportional to the position displacement. Please note for Gaussians with positional displacement magnitudes below $\epsilon$, the second term in \equref{eq:dif_scale} vanishes to $0$ and $\Delta G_{i,k}$ is always equal to $\Delta G^{init}_{i,k}$, which is inherently proportional to the positional displacement.

Instead of optimizing $\Delta G_{i,k}$, we directly optimize $\Delta \widehat{G}_{i,k}$ using the loss functions described in the following section. Specifically, $\Delta \widehat{G}_{i,k}$ is initialized to 0. Each time $\Delta \widehat{G}_{i,k}$ is updated, we calculate $\Delta G_{i,k}$ according to \equref{eq:dif_scale}, from which the avatar model is constructed. The avatar model is then rendered to an image through Gaussian splatting, which is used in loss function computation.

In this way, we effectively guide the Gaussian differences to change consistently with positional displacements, leading to optimized Gaussian blendshapes with strong semantic consistency with mesh blendshapes (see \figref{fig:prove_consistency}).

\subsection{Loss Functions}

The optimization goal is to minimize the image loss between the rendering and input, under some regularization constraints. The first loss is the image loss as in 3DGS~\cite{3DGS}, consisting of the $L_1$ differences between the rendered images and the video frames and a D-SSIM term:
\begin{equation}
L_{rgb} = (1-\lambda)L_1 + \lambda L_{D-SSIM} 
\end{equation}
with $\lambda = 0.2$.

We also design an alpha loss to constrain the Gaussians to stay within the head region.
We perform Gaussian splatting to get the accumulated opacity image $I_{\alpha}$, and compare it with the foreground head mask $Mask_h$. The alpha loss is defined as:
\begin{equation}
L_{\alpha} = \frac{1}{F}\sum_{i=1}^{F}(||(I_{\alpha}^i -Mask_{h}^i)||_2),
\end{equation}
where $F$ is the frame number.

We further introduce a regularization loss to constrain the mouth interior Gaussians to stay within a pre-defined volume of the mouth. Specifically, we compute the signed distance for each Gaussian to the volume boundary and apply an $L_2$ loss to retract it when it goes out of the volume. The regularization loss is defined as:
\begin{equation}
L_{reg} = \frac{1}{N}\sum_{i=1}^{N}(||max(SDF(\mathbf{x}_i,V),0)||^2_2),
\end{equation}
where $V$ is the pre-defined cylindrical volume, $\mathbf{x}_i$ is the Gaussian position and $N$ is the number of mouth interior Gaussians. The overall loss function is defined as:
\begin{equation}
L = \lambda_1L_{rgb}+\lambda_2L_{\alpha} + \lambda_3L_{reg}.
\end{equation}
We set $\lambda_1=1, \lambda_2=10, \lambda_3=100$ by default.

\begin{table}[t]
\caption{Quantitative comparisons between NeRFBlendShape~\cite{gao2022reconstructing} and our method.}
\label{tab:evaluation_nerfblendshape}
\centering
\tabcolsep=0.15cm
\resizebox{\linewidth}{!}{
\begin{tabular}{|cc|cccccc}
\hline
\multicolumn{2}{c|}{\multirow{2}{*}{Datasets}} & \multicolumn{6}{c}{NeRFBlendShape dataset} \\  \cline{3-8} \multicolumn{2}{c|}{} & id1 & id2 & id3 & id4 & id5 & id6  \\ \hline 
\multicolumn{1}{c|}{\multirow{3}{*}{PSNR $\uparrow$}}  & NeRFBlendShape   & 32.12  & 32.25  & 37.25 & 36.86 & 34.26  & 35.74 \\ 
\multicolumn{1}{c|}{}                       & Ours(w/o LPIPS)   & \cellcolor{c1}33.13 & \cellcolor{c1}33.23 & \cellcolor{c1}39.83 & \cellcolor{c1}38.34 & \cellcolor{c1}35.58 & \cellcolor{c1}36.59  \\ 
\multicolumn{1}{c|}{}                       & Ours(w/ LPIPS) & 33.09 & 33.15 & 39.64 & 38.17 & 35.46 & 36.46  \\ \cline{1-8}
\multicolumn{1}{c|}{\multirow{3}{*}{SSIM $\uparrow$}}  & NeRFBlendShape  & 0.9412 & 0.9369 & 0.9750 & 0.9791 & 0.9541 & 0.9786 \\ 
\multicolumn{1}{c|}{}                       & Ours(w/o LPIPS) & \cellcolor{c1}0.9532 & \cellcolor{c1}0.9473 & \cellcolor{c1}0.9836 & \cellcolor{c1}0.9846 & \cellcolor{c1}0.9635 & \cellcolor{c1}0.9814  \\ 
\multicolumn{1}{c|}{}                       & Ours(w/ LPIPS) & 0.9522 & 0.9457 & 0.9828 & 0.9837 & 0.9625 & 0.9806  \\ \cline{1-8}
\multicolumn{1}{c|}{\multirow{3}{*}{LPIPS $\downarrow$}} & NeRFBlendShape & 0.0715 & 0.0756 & 0.0436 & 0.0460 & 0.0448 & 0.0366  \\ 
\multicolumn{1}{c|}{}                       & Ours(w/o LPIPS) & 0.0862 & 0.0937 & 0.0486 & 0.0495 & 0.0538 & 0.0414 \\ 
\multicolumn{1}{c|}{}                       & Ours(w/ LPIPS) & \cellcolor{c1}0.0550 & \cellcolor{c1}0.0620 & \cellcolor{c1}0.0359 & \cellcolor{c1}0.0334 & \cellcolor{c1}0.0381 & \cellcolor{c1}0.0303  \\ \hline
\end{tabular}
}
\end{table}

\subsection{Implementation Details}
\label{sec:detail}

We implement our method using Pytorch. The Adam solver~\cite{Adam} is employed for parameter optimization. The learning rates are $3.2\times10^{-7}$, $5\times10^{-5}$, $5\times10^{-4}$, $1\times10^{-4}$, $1.25\times10^{-3}$ respectively for the Gaussian properties $\{\mathbf{x}_{k},\alpha_{k}, \mathbf{s}_{k}, \mathbf{q}_{k}, SH_{k}\}$. The initially sampled Gaussian number is 50k for the neutral model, and 14k for the mouth interior Gaussians.

The training is conducted on an A800 GPU and testing is conducted on an RTX 4090 GPU. We also build a C++/CUDA interactive viewer following 3DGS~\cite{3DGS} and use it to measure our runtime frame rates.

As Gaussian positions frequently change during optimization, we need to efficiently update their LBS blend weights $\mathbf{w}$ and the positional displacements of nearest points $\{d_{i,k}\}$. We precompute and store these values on a 3D grid of $256\times256\times256$ surrounding the neutral mesh $M_0$. The values of an arbitrary Gaussian can be effectively computed as the linear blending of the values of eight grid points nearest to the Gaussian center.

\begin{figure}[t]
\centering
\makebox[\linewidth]{\includegraphics[width=\linewidth]{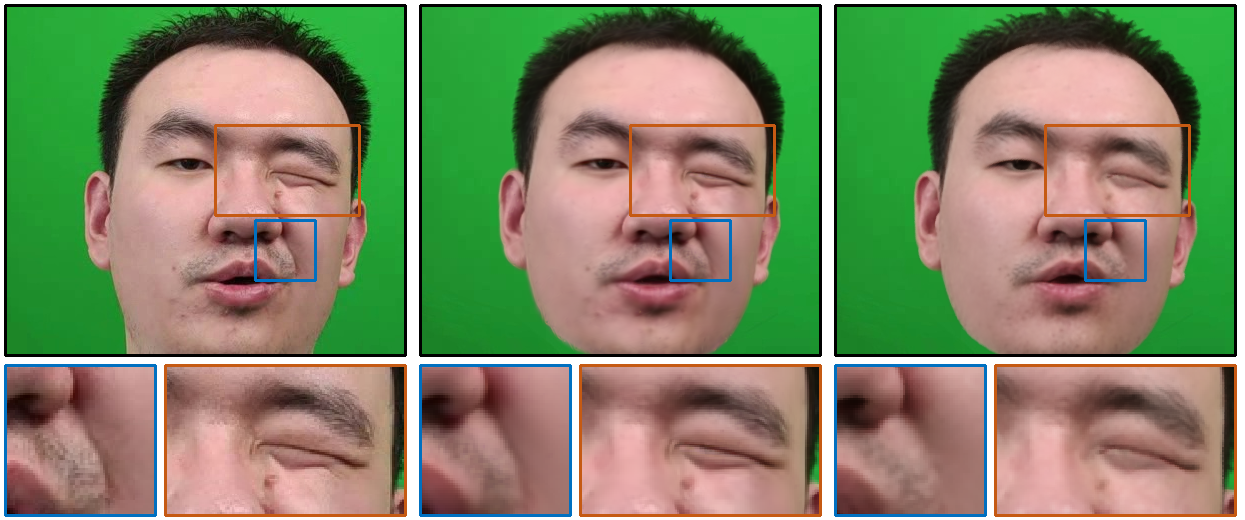}}
\makebox[\linewidth]{\includegraphics[width=\linewidth]{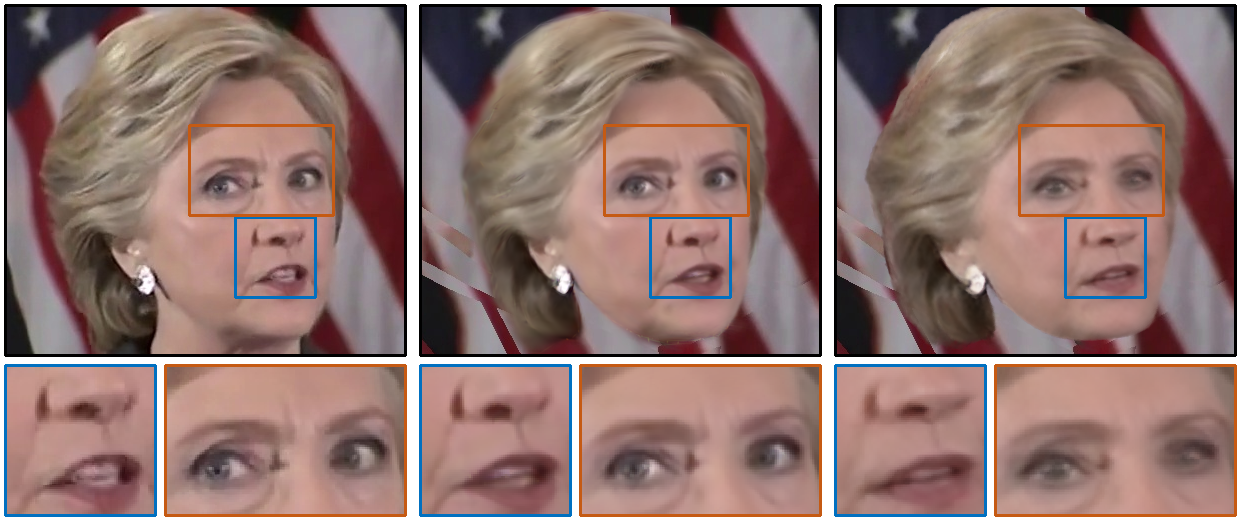}}
\makebox[0.32\linewidth]{Ground truth}
\makebox[0.32\linewidth]{Ours}
\makebox[0.32\linewidth]{NeRFBlendShape} \\
\caption{Qualitative comparisons with NeRFBlendShape~\cite{gao2022reconstructing}. Our method more faithfully captures fine facial details (e.g., wrinkles around the eyes and nose), and better recovers the eyeball movement. YouTube video ID is -yHgE9W699w for Hillary Clinton.}
\Description{The first row shows a male with a skewed mouth expression, while the second row shows Hillary giving a speech.}
\label{fig:compare_nerfbs}
\end{figure}

\begin{figure}[h]
    \setlength{\columnsep}{0pt}
   \begin{minipage}[c]{\dimexpr0.03\linewidth}
    \rotatebox{90}{
    \begin{tabular}{p{2.2cm}p{2.0cm}p{1.3cm}}
        PointAvatar & INSTA & Ours
    \end{tabular}
    }
  \end{minipage}
  \begin{minipage}[c]{\dimexpr0.95\linewidth-\columnsep}
    \includegraphics[width=\linewidth]{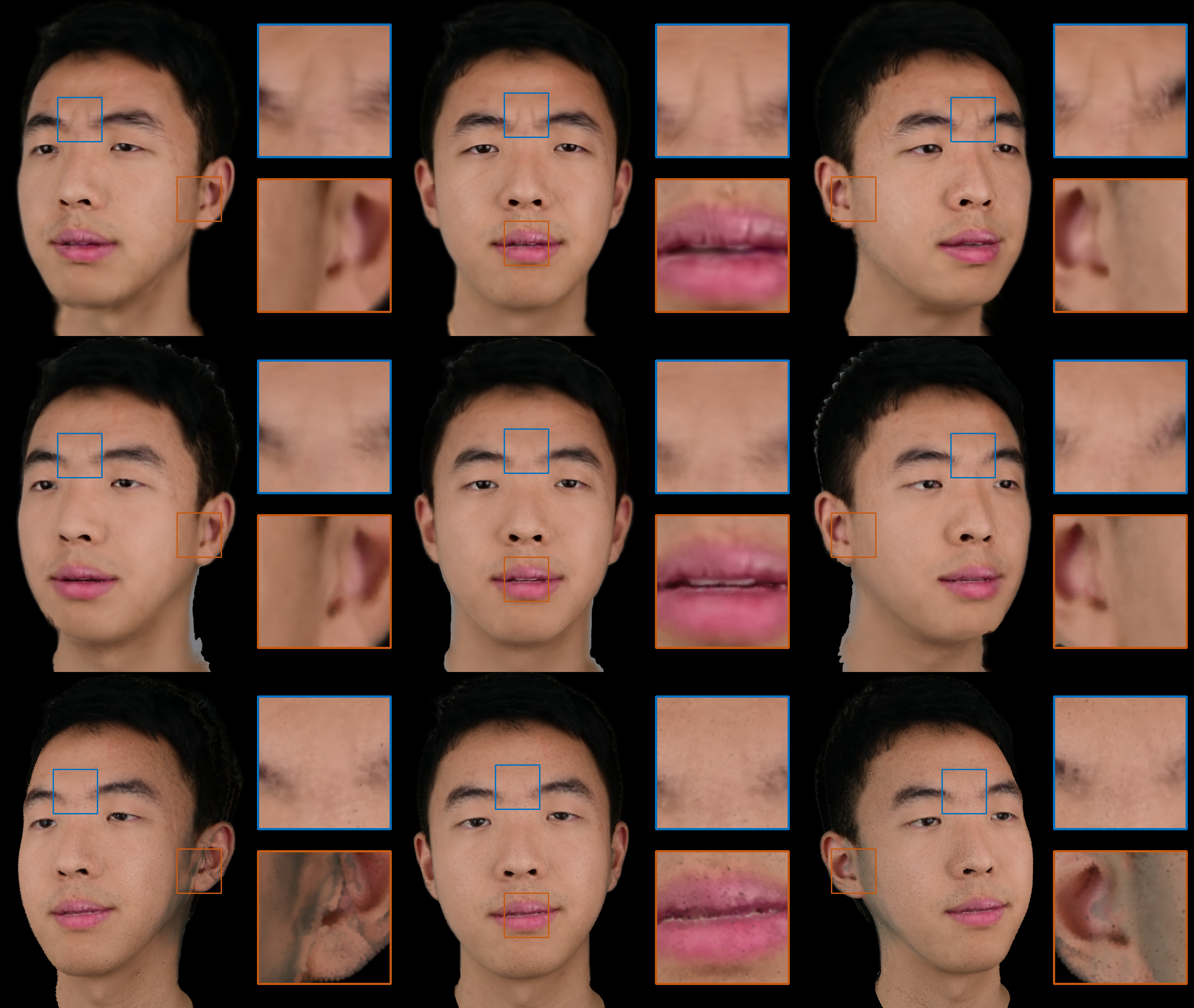}
  \end{minipage}
\caption{Qualitative comparisons for novel view extrapolation. Our method produces better results with fine details under novel views.}
\Description{An avatar with a frown expression is displayed from the left, front, and right perspectives.}
\label{fig:compare_novel_view}
\end{figure}

\section{Results}

\subsection{Baselines and Datasets}

We compare our method with state-of-the-art methods, NeRF-based INSTA~\cite{zielonka2023instant} and point-based PointAvatar~\cite{zheng2023pointavatar}, on the INSTA dataset and our own dataset. We hold the last 350 frames of each video as the test set for self-reenactment similar to INSTA. The training data preparation time is about 12 hours for 4500 frames as in INSTA. We also compare our method with NeRFBlendShape~\cite{gao2022reconstructing} on their public dataset consisting of eight videos. The last 500 frames of each video are reserved for test. We reduce the alpha weight $\lambda_2$ to 1 for the entire dataset due to the relatively inaccurate binary foreground mask provided, which we find slightly sharpens the hair around the contour.

Our own dataset consists of four subjects, captured in an indoor environment using a Nikon D850 camera. For each subject, we collected a 3 minute video in 1080p, which was then cropped and resized to $1024 \times 1024$ resolution.

\begin{figure}[t]
\setlength{\columnsep}{0pt}
  \begin{minipage}[c]{\dimexpr0.03\linewidth}
    \rotatebox{90}{
    \begin{tabular}{p{2.2cm}p{1.5cm}}
        Transferred & Source
    \end{tabular}
    }
  \end{minipage}
  \begin{minipage}[c]{\dimexpr0.95\linewidth-\columnsep}
    \includegraphics[width=\linewidth]{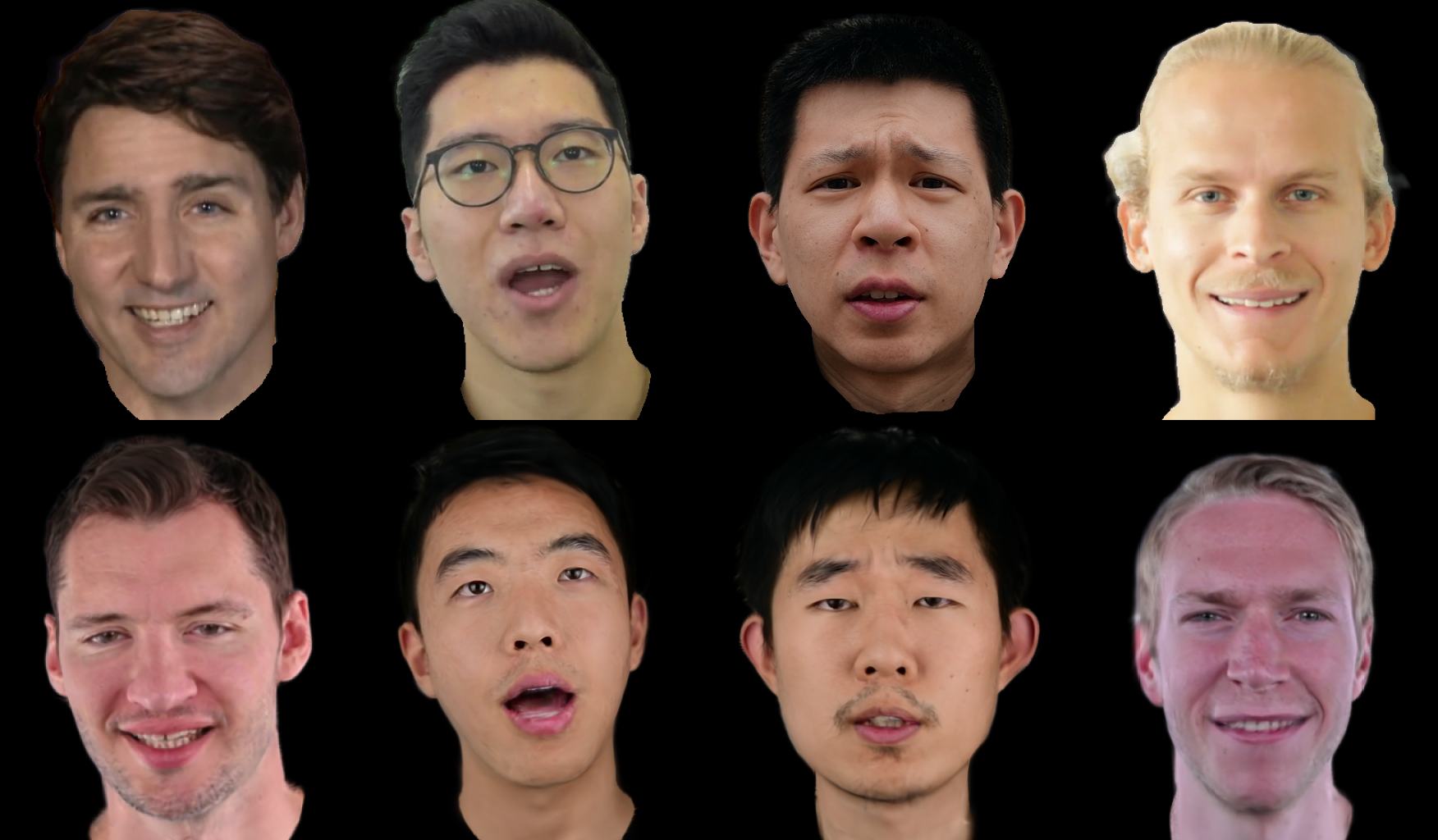}
  \end{minipage}
\caption{Results of cross-identity reenactment. YouTube video ID is mKHgXHKbJUE for Justin Trudeau.}
\Description{The transfer of expressions from one individual to another is demonstrated, with four examples including smiling, opening the mouth, frowning and grinning.}
\label{fig:cross_identity}
\end{figure}

\begin{figure}[t]
\centering
\includegraphics[width=0.99\linewidth]{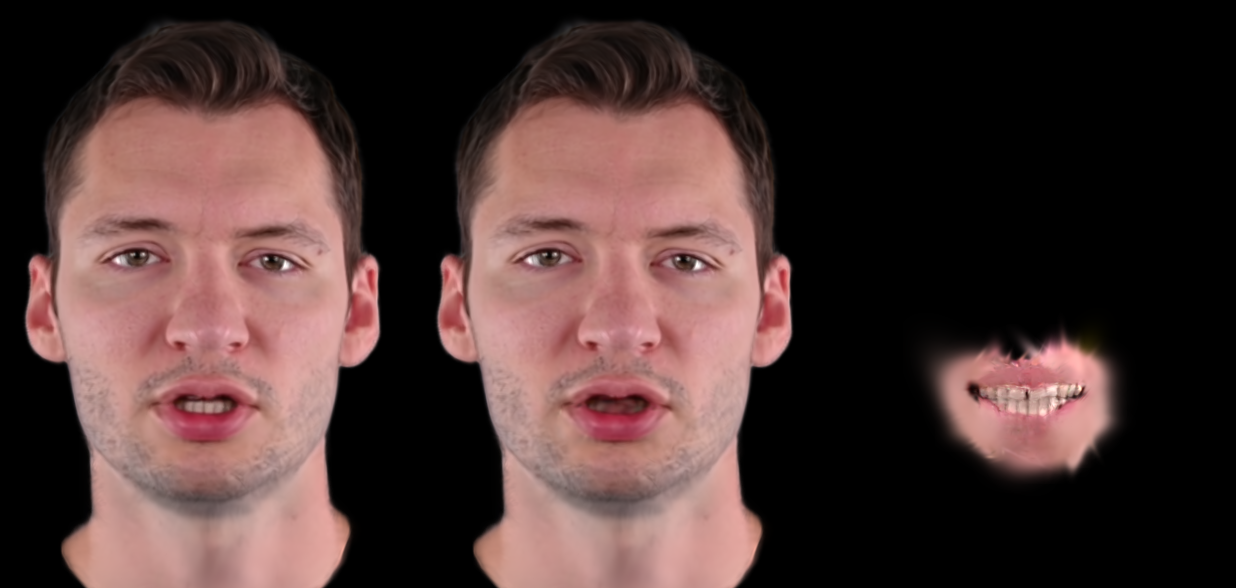}
\makebox[0.32\linewidth]{{Full model}}
\makebox[0.32\linewidth]{{w/o mouth Gaussians}}
\makebox[0.32\linewidth]{{mouth Gaussians}} \\
\caption{Demonstration of mouth interior Gaussians.}
\Description{The image displays the appearance of our model after being segmented into the mouth interior Gaussians and the remaining parts.}
\label{fig:vis_component}
\end{figure}

\begin{table}[t]
\caption{Performance comparisons. We perform training for all methods on an A800 GPU. Testing is done on a RTX 4090 GPU for INSTA, NeRFBlendshape, and our method, but on an A800 GPU for PointAvatar due to out-of-memory errors on the RTX 4090 GPU. The rendering resolution is $512 \times 512$. Note our running time includes both animation (i.e., linear blending and LBS transformation) and rendering, and our performance is insensitive to the rendering resolution. We also report the peak memory consumption during training and runtime computation.}
\label{tab:performance}
\centering
\tabcolsep=0.15cm
\resizebox{\linewidth}{!}{
\begin{tabular}{l|rrrc}
\toprule
& Training & Runtime & Mem. (train) &  Mem. (runtime) \\ 
\midrule
INSTA & 10min & 70fps & 16G & 4G \\
 PointAvatar & 3.5h & {5fps}$^{\ast}$ & 40G &  {32G}$^{\ast}$ \\
NeRFBlendShape & 20min & 26fps & 7G &  2G \\
Ours & 25min & 370fps  & 14G &  2G \\
\bottomrule
\end{tabular}
}
\end{table}

\subsection{Comparisons}

We evaluate the results using standard metrics including PSNR, SSIM and LPIPS~\cite{zhang2018unreasonable}. As show in the quantitative results~\tabref{tab:evaluation_INSTA}, in most cases, our method outperforms INSTA and pointAvatar in terms of PSNR and LPIPS, while the SSIM of our method is consistently better.
As shown in~\tabref{tab:evaluation_nerfblendshape}, our method also surpasses NeRFBlendShape in terms of PSNR and SSIM. Note that NeRFBlendShape utilizes the LPIPS loss during training, leading to better LPIPS. When we add the LPIPS loss with a weight of $0.05$ in training, our method also performs better. 

\begin{figure*}[h]
\centering
\makebox[\linewidth]
{\includegraphics[width=\linewidth]{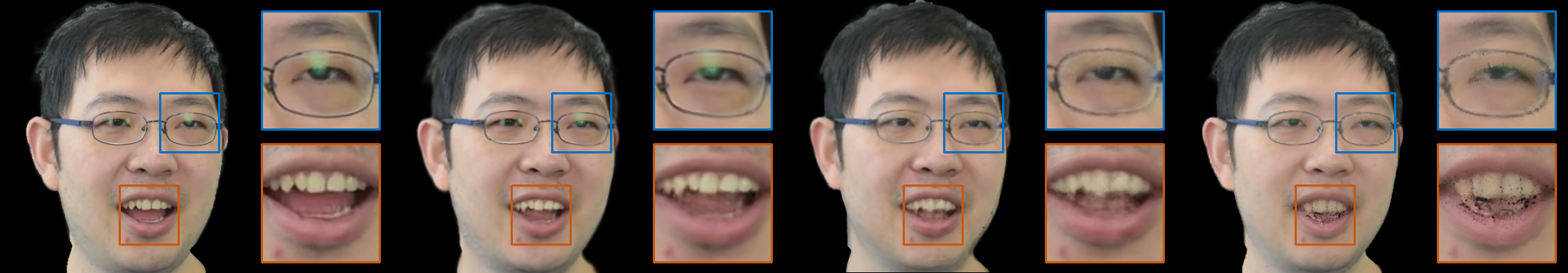}}
\makebox[\linewidth]
{\includegraphics[width=\linewidth]{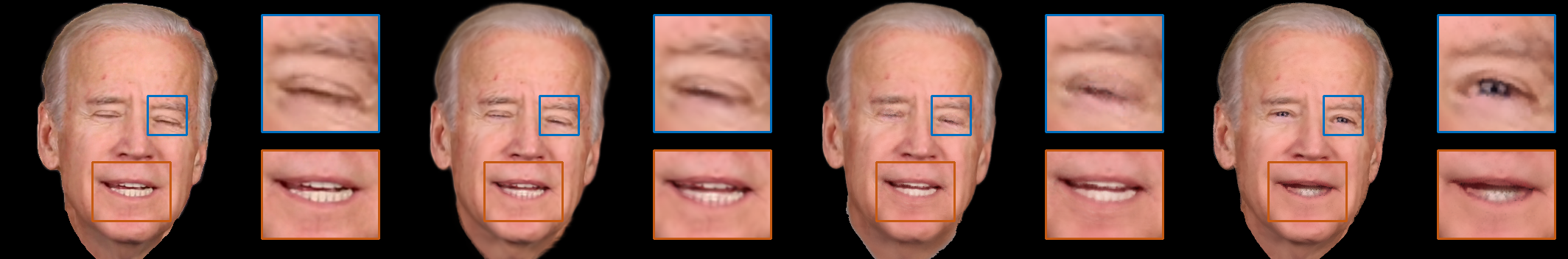}}
\makebox[\linewidth]{\includegraphics[width=\linewidth]{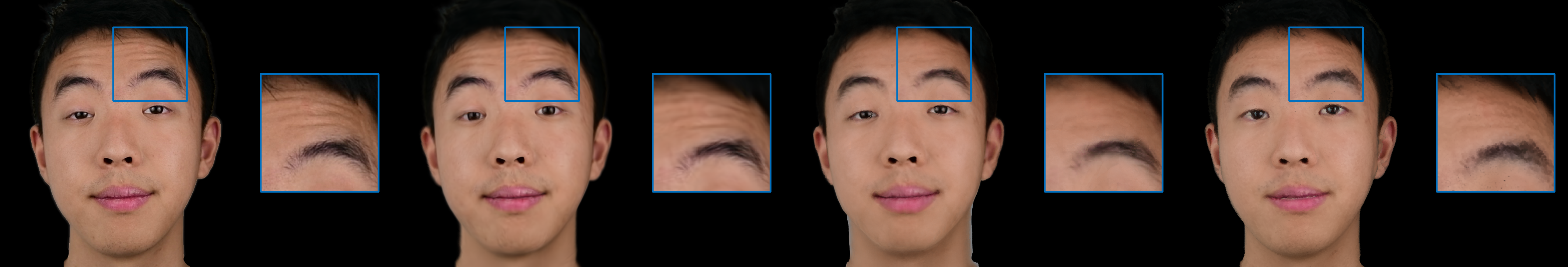}}
\makebox[\linewidth]{\includegraphics[width=\linewidth]{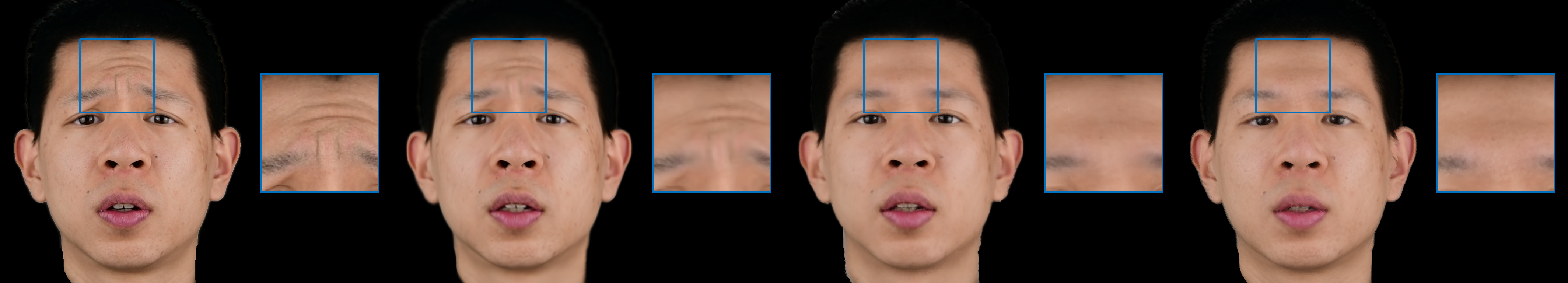}}
\makebox[0.24\linewidth]{Ground truth}
\makebox[0.24\linewidth]{Ours}
\makebox[0.24\linewidth]{INSTA}
\makebox[0.24\linewidth]{PointAvatar}
\caption{Qualitative comparisons between INSTA~\cite{zielonka2023instant}, PointAvatar~\cite{zheng2023pointavatar}, and our method. Our method better captures high-frequency details and specular highlights. YouTube video ID is smghyezLW5o for Joe Biden.}
\Description{From top to bottom, the images respectively show a male talking with an open mouth, Biden with closed eyes, a male with raised eyebrows, and a male with a frown expression.}
\label{fig:compare_INSTA}
\end{figure*}

The qualitative comparisons are shown in \figref{fig:compare_INSTA} and \figref{fig:compare_nerfbs}. 
Compared with INSTA and PointAvatar, our method is better at capturing high-frequency details observed in the training video, such as wrinkles, teeth, and specular highlights of glasses and noses.
Compared with NeRFBlendShape, our method also synthesizes images of higher quality with sharper details. Moreover, our method better recovers the eyeball movement than NeRFBlendShape, thanks to the eyeball motion control provided in FLAME.

\begin{figure}[t]
\centering
\includegraphics[width=0.32\linewidth]{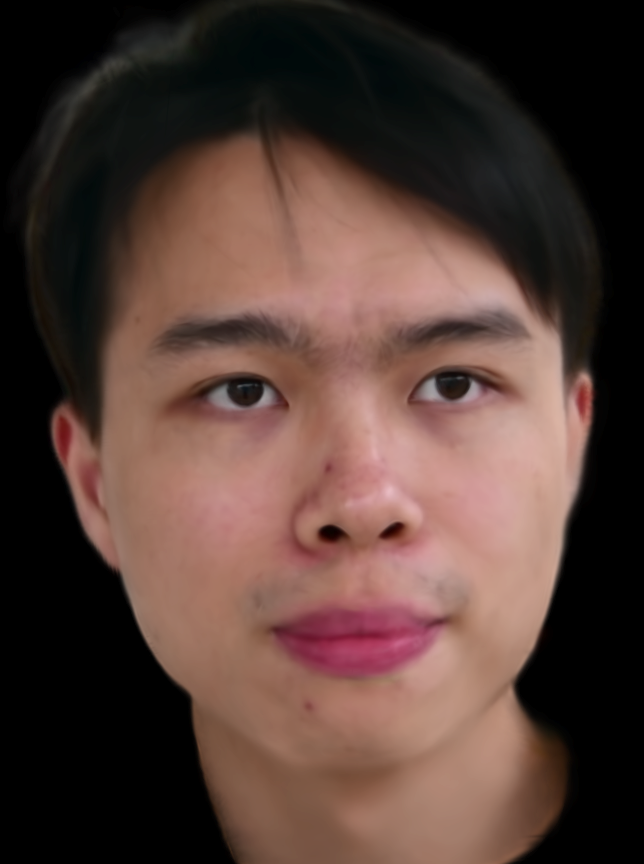}
\includegraphics[width=0.32\linewidth]{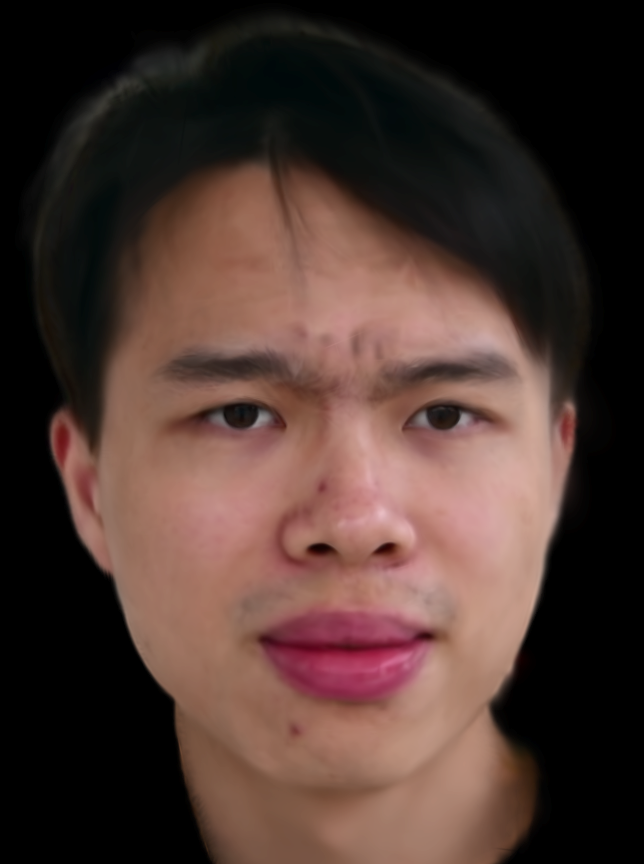}
\includegraphics[width=0.32\linewidth]{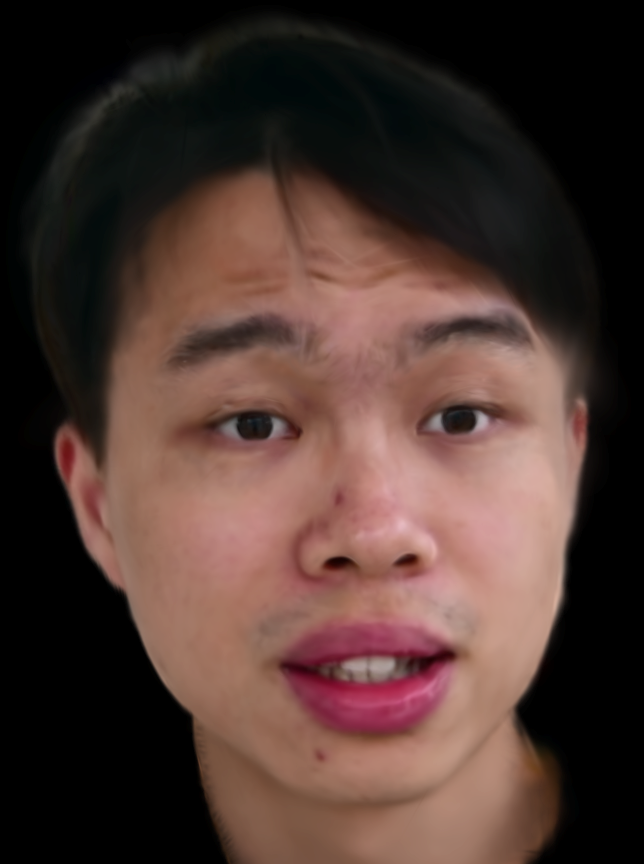}
\includegraphics[width=0.32\linewidth]{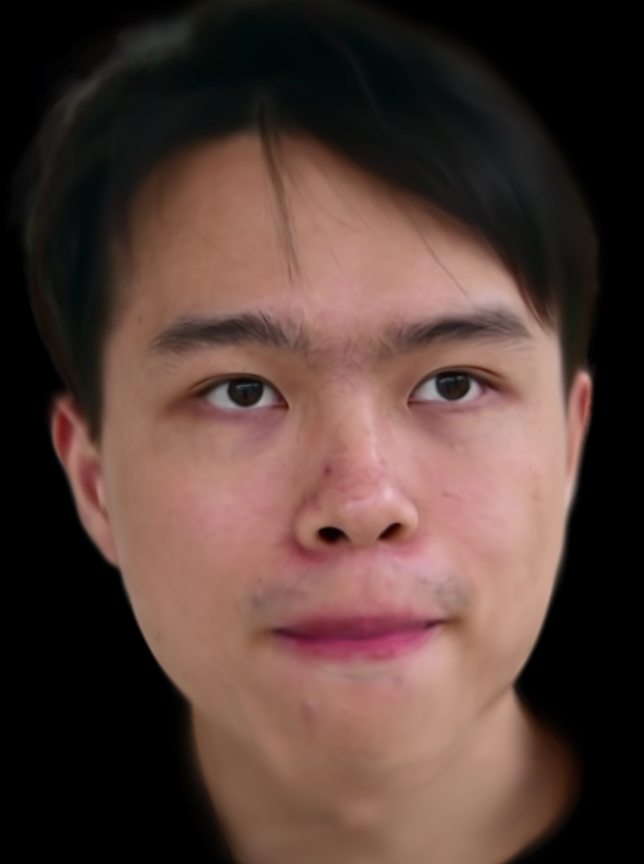}
\includegraphics[width=0.32\linewidth]{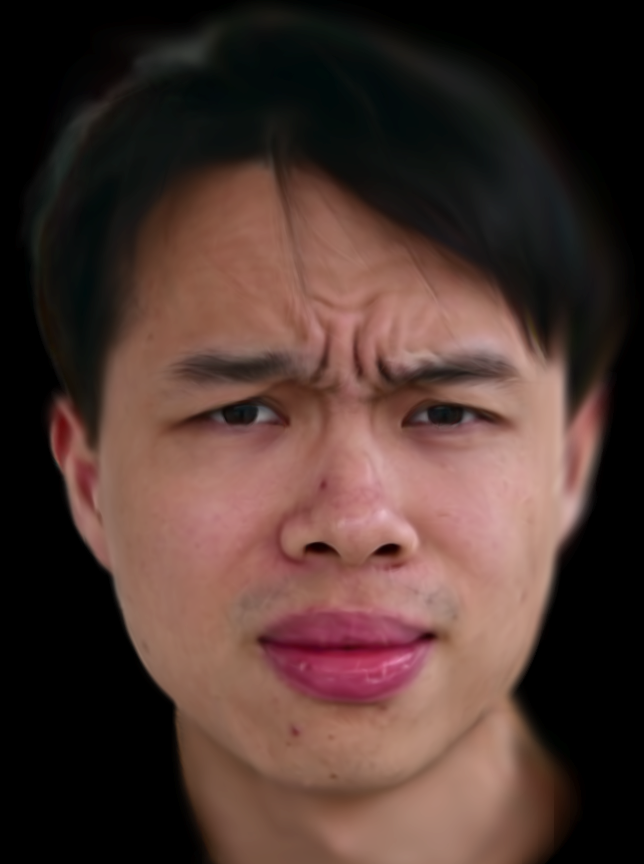}
\includegraphics[width=0.32\linewidth]{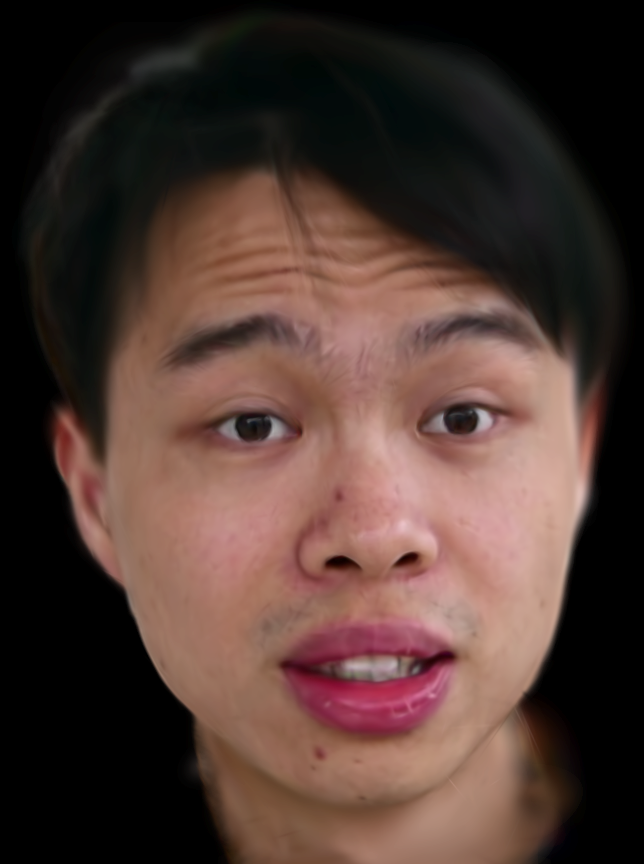}
\caption{Ablation study on blendshape optimization. The first row shows the results with the initial values of $\{\Delta B_k\}$ kept unchanged during optimization. The second row shows our results with joint optimization of $B_0$, $\{\Delta B_k\}$, and $B_m$, which better capture intricate details of facial animations.}
\Description{The three expressions shown in the image are pursing lips, frowning and raising eyebrows. In the second row, the action of pursing the lips is more pronounced, and the wrinkles are also more noticeable.}
\label{fig:ablation_blendshape_opt}
\end{figure}

Our method also performs better in novel view extrapolation (\figref{fig:compare_novel_view}), while PointAvatar~\cite{zheng2023pointavatar} suffers from artifacts around the ear region, and both INSTA~\cite{zielonka2023instant} and PointAvatar tend to lose high-frequency details.

We show qualitative results on cross-identity reenactment in \figref{fig:cross_identity}. Our method faithfully transfers expressions while maintaining the personal attributes of the target subject. 

The training and runtime performance comparison is shown in~\tabref{tab:performance}. Our method is able to synthesize facial animations at 370fps, over $5\times$ faster than INSTA and about $14\times$ faster than NeRFBlendshape. Our training time is comparable to NeRFBlendshape.

\begin{figure}[t]
\centering
\makebox[\linewidth]{\includegraphics[width=\linewidth]{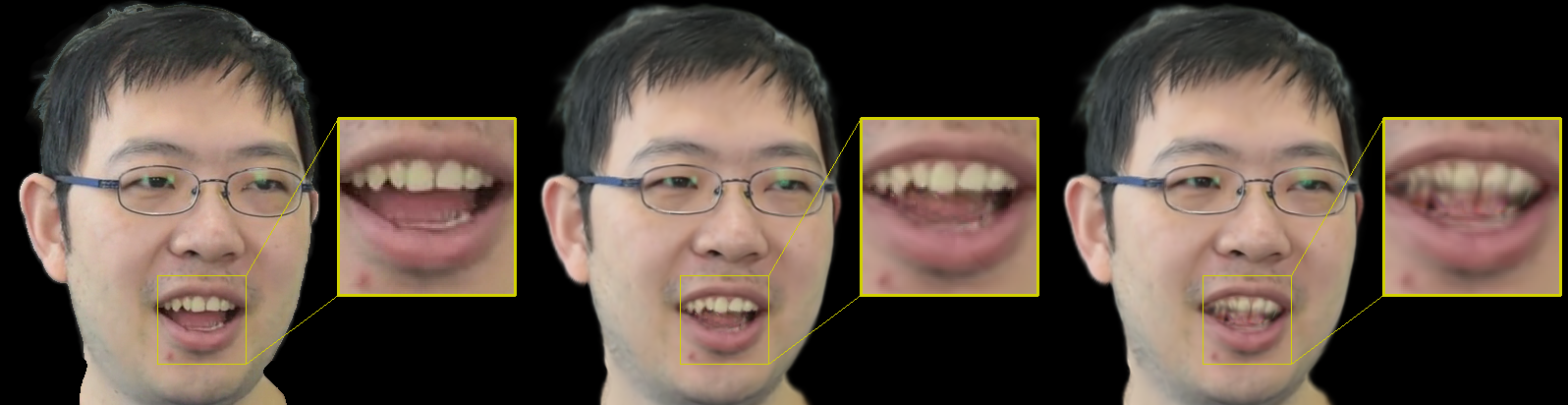}}
\makebox[0.3\linewidth]{Ground truth}
\makebox[0.3\linewidth]{Ours}
\makebox[0.3\linewidth]{w/o mouth Gaussians}
\caption{Ablation study on mouth interior Gaussians. We find that without the mouth interior Gaussians, the teeth may not be well modeled, leading to blurry or ghosting artifacts. }
\Description{Figure 10. Fully described in the text.}
\label{fig:ablation_mouth_hair}
\end{figure}

\begin{figure}[h]
\centering
\makebox[\linewidth]{\includegraphics[width=\linewidth]{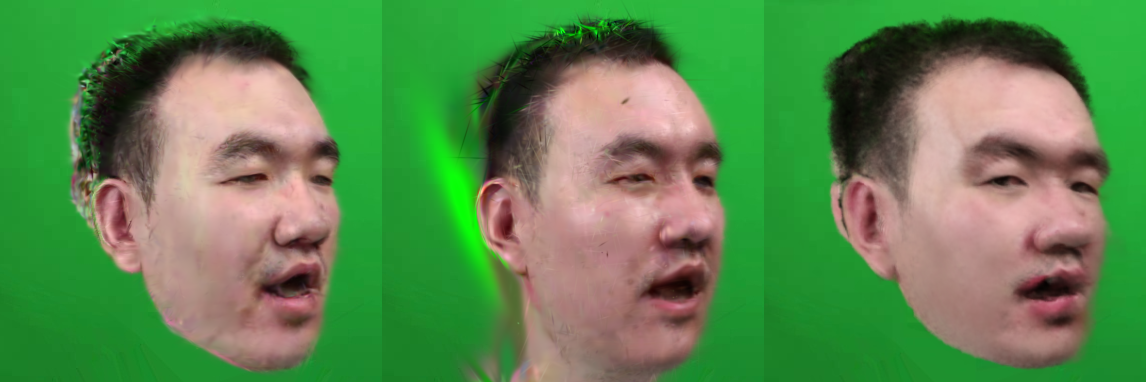}}
\makebox[0.3\linewidth]{Ours}
\makebox[0.3\linewidth]{FlashAvatar}
\makebox[0.3\linewidth]{NeRFBlendShape}
\caption{Failure cases of side view rendering.}
\Description{The results of an extreme side view for our model, FlashAvatar and NeRFBlendShape are displayed. Artifacts are obvious in all three methods.}
\label{fig:side_view}
\end{figure}

\subsection{Gaussian Blendshape Visualization}
\begin{figure*}[p]
\centering
\includegraphics[width=0.99\linewidth]{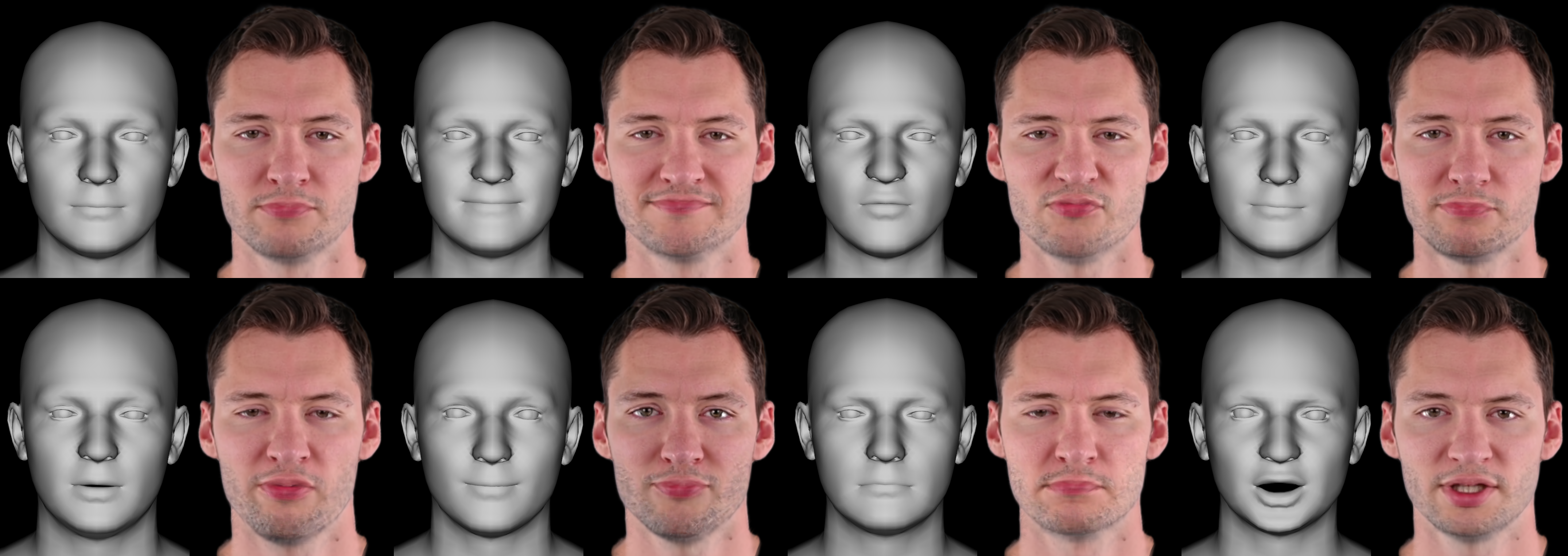}
\caption{Visualization of our Gaussian blendshapes. Each Gaussian blendshape resembles its corresponding FLAME mesh blendshape, and captures photo-realistic appearance.}
\Description{Eight examples are given, each showing a Gaussian blendshape and its corresponding mesh blendshape.}
\label{fig:vis_blendshape}
\end{figure*}

\figref{fig:vis_blendshape} demonstrates eight Gaussian blendshapes of a subject and their corresponding mesh blendshapes. Please refer to the supplementary video for live demonstration. The effect of our mouth Gaussians is show in \figref{fig:vis_component}. 

\subsection{Ablation Studies}

\paragraph{Blendshape consistency.} 
\figref{fig:prove_consistency} visualizes the magnitudes of $\Delta M_k$ and $\Delta B_k$. As you can see, imposing blendshape consistency during optimization does produce Gaussian blenshapes $\{B_k\}$ differing from the base model $B_0$ in a consistent way that mesh blendshapes $\{M_k\}$ differ from the base mesh $M_0$. \figref{fig:ablation_artifact} demonstrates the importance of blendshape consistency between Gaussian and mesh blendshapes. The optimization without considering blendshape consistency on all Gaussian properties results in apparent artifacts on the face and head boundary under novel expressions.
Note that the magnitude of $\Delta B_k$ visualized in \figref{fig:prove_consistency} only represents the difference between each individual blendshape $B_k$ and the base model $B_0$, and thus does not necessarily correspond to errors in rendered images of the avatar model, which is the linear blending of all blendshapes.

\paragraph{Optimization of $\{\Delta B_k\}$.}
The initialization stage of our training constructs Gaussian blendshapes $\{B_k\}$ by transforming Gaussians of $B_0$ using the deformation gradients from $M_0$ to $\{M_k\}$, resulting in Gaussian differences $\{\Delta B_k\}$ consistent with mesh differences $\{\Delta M_k\}$. Keeping the initial values of $\{\Delta B_k\}$ unchanged during optimization and only optimizing $B_0$ and $B_m$ can produce reasonable results, but fail to capture the fine details of facial animations, as shown in \figref{fig:ablation_blendshape_opt}.

\paragraph{Mouth interior Gaussians}
We evaluate the effect of mouth interior Gaussians by comparing our full result with the one using only the neutral model and expression blendshapes to represent the whole head (\figref{fig:ablation_mouth_hair}). We can see that apparent artifacts and blurring occur around the mouth region, demonstrating the necessity of mouth interior Gaussians.

\section{Conclusion}

We present a novel 3D Gaussian blendshape representation for animating photorealistic head avatars. We also introduce an optimization process to learn the Gaussian blendshapes from a monocular video, which are semantically consistent with their corresponding mesh blendshapes. Our method outperforms state-of-the-art NeRF and point based methods in producing avatar animations of superior quality at significantly faster speeds, while the training and memory cost is moderate.

\paragraph{Limitation and Discussion.} Our constructed avatar models can exhibit apparent artifacts in side-view rendering if the training data does not contain side views. As shown in \figref{fig:side_view}, this is also a limitation in previous NeRF-based methods and concurrent Gaussian-based methods. Improving the generalization capability to handle dramatically novel views is an open problem for further research. The extrapolation capabilities of our model are also restricted by its linear blending nature of the model, leading to potential failures when processing exaggerated expressions unseen in the training set. Another limitation is that the model cannot represent deformable hair, which is an interesting direction for future investigation.
It is worth noting that there is a risk of misuse of our method (e.g., the so-called DeepFakes). We strongly oppose applying our work to produce fake images or videos of individuals with the intention of spreading false information or damaging their reputations.

\begin{acks}
This work is supported by the National Key Research and Development Program of China (No. 2022YFF0902302), NSF China (No. 62172357 \& 62322209), and the XPLORER PRIZE. The source code is available at \url{https://gapszju.github.io/GaussianBlendshape}.
\end{acks}

\bibliographystyle{ACM-Reference-Format}
\bibliography{sample-base}

\end{document}